\documentclass[]{article} \usepackage{aaspp4}  
\usepackage{graphicx}
\usepackage{natbib}

\begin{document}
\title{\bf Persistent Evidence of a Jovian Mass Solar Companion in the Oort Cloud}

\author{John J. Matese and Daniel P. Whitmire \\Department of Physics\\
University of Louisiana at Lafayette, Lafayette, LA, 70504-4210 USA}

\begin{center}
\vspace{1cm} Pages: 41 including 2 cover sheets, 39 text pages, 2 tables, 9 figures\\
\vspace{1cm} Submitted to the journal ICARUS on 15 April 2010.\\
\end{center}
\newpage
{\bf PROPOSED RUNNING HEAD}\\
``Jovian Mass Solar Companion in the Oort Cloud"\\
\vspace{1cm}

{\bf EDITORIAL CORRESPONDENCE AND PROOFS}\\
John J. Matese\\
Department of Physics\\
University of Louisiana at Lafayette\\
Lafayette, LA, 70504-4210 \\
Tel: (337) 482 6697\\
Fax: (337) 482 6699\\
E-mail: matese@louisiana.edu\\
\newpage
\begin{quotation}
{\bf  Abstract:}
We present an updated dynamical and statistical analysis of outer Oort cloud cometary
evidence suggesting the sun has a wide-binary Jovian mass companion. The
results support a conjecture that there exists a companion of mass $\approx1-4\;{\rm M_{Jupiter}}$ orbiting in the innermost region of the outer Oort cloud. Our most restrictive prediction is that the orientation angles of the orbit normal in galactic
coordinates are centered on $\Omega$,  the galactic longitude of the
ascending node = 319$^\circ$ and $i$, the galactic inclination =
103$^\circ$ (or the opposite direction) with an uncertainty in the
normal direction subtending $\approx 2\%$ of the sky. A Bayesian
statistical analysis suggests that the probability of the
companion hypothesis is comparable to or greater than the
probability of the null hypothesis of a statistical fluke. Such
a companion could also have produced the detached Kuiper Belt
object Sedna. The putative companion could be easily detected by the recently launched
Wide-field Infrared Survey Explorer (WISE).
\end{quotation}
\vspace{1cm} {\em Key Words:} Comets, dynamics; Celestial
mechanics; Kuiper Belt; Jovian planets; Planetary dynamics
\newpage

\section{Introduction}
Anomalies in the aphelia distribution and orbital elements of
Outer Oort cloud comets led to the suggestion that $\approx$ 20\%
of these comets were made discernable due to a weak impulse from
a bound Jovian mass body  (\cite{mww99}). Since that time the
data base of comets has doubled. Further motivation for an
updated analysis comes from the recent launch of the Wide-field
Infrared Survey Explorer (WISE; \cite{wise}), which could easily
detect the  putative companion orbiting in
the outer Oort cloud. Such an object would be incapable of
creating comet ``storms". To help mitigate popular confusion with
the Nemesis model (\cite{wj84}, \cite{dhm84}) we use the name
recently suggested by Kirkpatrick and Wright (2010), Tyche, (the
good sister of Nemesis) for the putative companion.

The outer Oort cloud (OOC) is formally defined as the ensemble of
comets having original semimajor axes $A \geq 10^4$ AU
(\cite{oort50}). It has been shown that the majority of these
comets that are made discernable are first-time entrants into the
inner planetary region (\cite{fern81}) and these comets are
therefore commonly referred to as {\em new}. The dominance of the
galactic tide in making OOC comets discernable at the present
epoch has been predicted on theoretical grounds (\cite{ht86}).
Observational evidence of this dominance has been claimed to be
compelling (\cite{del87,mw92,wt99,ml04}).

Matese and Lissauer (2004) adopted an  {\em in situ} energy
distribution similar to the initial distribution of Rickman {\em
et al.} (2008) and took the remaining phase space external to a
``loss cylinder"  to be uniformly populated at the present epoch.  The distribution
of cometary orbital elements made discernable from the tide alone
was then obtained and compared with observations.

Similar modeling (\cite{ml02}) had been performed including single stellar impulses
which mapped the comet flux over a time interval of 5 Myr,
in 0.1 Myr intervals. Peak impulsive enhancements $\ge 20\%$
were found to have a half-maximum duration of $\approx 2$ Myr and occurred
with a mean time interval of $\approx 15 $ Myr. Various
time-varying distributions of elements were compared with the
modeled tide-alone results  and inferences about the
signatures of a weak stellar impulse were drawn.

In Section 2 we review  a discussion (\cite{ml04}) of a subtle characteristic of
galactic tidal dominance which is difficult to mimic with
observational selection effects or bad data. Along with the more
well known feature of the deficiency of major axis orientations
in the direction of the galactic poles and equator, we compare
with observations these predictions based on the tidal
interaction alone and show that the data are of sufficiently high
quality to unambiguously demonstrate the dominance of the
galactic tide in making comets discernable at the present epoch.
A critique of objections to this assertion (\cite{rffv08}) is
also presented. More recent detailed modeling (\cite{kaib09})
provide important insights into the evolving populations of the
{\em in situ} and discernable populations of the Oort cloud. We
comment further on these works in this section.

In Section 3 we describe the theoretical analysis combining a
secular approximation for the galactic tide and for a point mass
perturber, describing how a {\it weak} perturbation
of OOC  comets would manifest itself observationally. Evidence
suggesting that there is such an aligned impulsive component of
the observed OOC comet flux has been previously reported
(\cite{mww99}). It has been found that none of the known
observational biases can explain the alignment found there
(\cite{he02}). The size of the available data  has since doubled
which leads us to review the arguments here.

In Section 4 we present the supportive evidence that an impulsive
enhancement in the new comet flux of $\approx 20\%$ persists in
the updated data. We also discuss dynamical and observational
limits on parameters describing the putative companion. Section 5 summarizes our results and presents our conclusions.

\section{Secular dynamics of the galactic tide}
Near-parabolic comets are most likely to have their perihelia
reduced to the discernable region. The dynamics of the galactic
tide acting on near-parabolic OOC comets is most simply described
in a Newtonian framework (\cite{mww99}, \cite{ml02}, \cite{ml04}).
A summary of their analyses is now given and followed
with the evidence that the galactic tidal perturbation
dominates in making OOC comets discernable at the present epoch.

\subsection{Theory}
Saturn and Jupiter provide an effective
dynamical barrier to the migration of OOC comet perihelia. OOC
comets that are approaching the planetary zone at the present
time were unlikely to have had a prior perihelion, $q_{prior}$,
that was interior to the ``loss cylinder" radius, $q_{lc}\approx
15$ AU, when it left the planetary region on the present orbit.
The simplifying assumption $q_{lc}\leq q_{prior}$ is then made for
the present orbit. During the present orbit comet perihelion will
then have been changed by the galactic tide (and by any
putative companion or stellar perturbation). The orbital
elements just before re-entering the planetary region on the
present orbit are commonly referred to as ``original" and will
be, in essence, the observed values with the exception of the
semimajor axis (perturbations by the major planets do not
significantly change any other orbital element of OOC comets).
Thus $q_{prior}$ is changed to $q_{original} \approx q_{obs}$,
the observed value, during the course of the present orbit. As an
observed comet comes  within a discernable region ($q_{obs} \leq
q_{discernable}\approx 5$AU)  and leaves the planetary region
again, the semimajor axis will have been changed from
$A_{original}$ to $A_{future}$, {\em i.e.}, $A_{future}\equiv
A_{prior}$ for the {\em next}  orbit. The comet is most likely
ejected  or turned into an inner Oort cloud comet with a small
fraction returning as OOC comets. Daughter comets returning to the discernable zone are likely to have faded and be more difficult to observe (\cite{wt99}). The $17th$  Catalogue of Cometary Orbits (\cite{mars08}) indicates that $\approx 14\%$ of observed
original OOC comets exit the planetary region as future OOC
comets. Therefore the discernable population of OOC comets should
be dominated by first time entrants to the loss cylinder. In the
following  we adopt the notation $ A\equiv A_{original}$.

For near-parabolic comets, the angular momentum per unit mass
determines the perihelion distance, $\mathbf{ H}\equiv \mathbf{ R}\times
{\dot\mathbf{ R}}$, ($\mathbf{ H}\perp\mathbf{ q},\; H\approx\sqrt{2\mu_\odot
q}$). With these assumptions, an observed OOC comet entering the
loss cylinder region for the first time had perihelion distances
$q_{obs}\leq q_{discernable} < q_{lc}\leq q_{prior}$. Therefore,
reducing $q$ in a single orbit requires a decrease in angular
momentum from the galactic tidal torque (and/or from angular
momentum changes by the putative companion or star),
\begin{equation}
\Delta \mathbf{H} \equiv \mathbf{H_{obs}} - \mathbf{H_{prior}} \;
 \textrm{, or,} \; \;\; q_{prior} - q_{obs} =  \left( \mathbf{\Delta H}^2 -  2 \mathbf{ H_{obs}} \cdot \mathbf{\Delta H}  \right)/2{\mu_\odot}.
\label{eqs:DeltaH}
\end{equation}
The weakest perturbation that could make a comet discernable
would reduce the prior perihelion distance from $q_{prior} \approx
q_{lc}$ to $q_{obs} \approx q_{discernable}$ (see Fig. 1) such that
\begin{equation}
 {|\Delta \mathbf{H}|}_{min} = \sqrt {2\mu_\odot \; q_{lc}}-\sqrt {2\mu_\odot \;  q_{discernable}}\;\;.
\label{eqs:DeltaHmin}
\end{equation}
Also of interest is the evolution of the aphelion orientation, ${\mathbf{\hat Q}}\equiv ~ ({\cos B}\;{\cos L}, {\cos B}\;{\sin L }, \sin B)$, expressed in terms of the aphelion latitude, $B$, and longitude, $L$.

If the galactic tide dominates in making OOC comets discernable we can recast Eq. (1)  as
$
q_{prior} - q_{obs} =  \left( \mathbf{\Delta H_{tide}}^2 -  2 \mathbf{ H_{obs}} \cdot \mathbf{\Delta H_{tide}}  \right)/2{\mu_\odot}
$
from which it is implied that if {\it weak} tidal perturbations
dominate in making OOC comets discernable, the tidal
characteristic,
$
S\equiv {\rm Sign}(\mathbf{ H_{obs}}  \cdot \Delta \mathbf{ H_{tide}})
$
will more often be $\;-1$ than +1.  Detailed modeling
results (\cite{ml04}) confirm this implication. This
characteristic combination of observed orbital elements forms an
{\em essential aspect of the present analysis} and has not been
included in modeling results presented elsewhere ({\em e.g.},
\cite{rffv08}, \cite{kaib09}).


\begin{figure}[!hb] \includegraphics[width=.75 \textwidth]{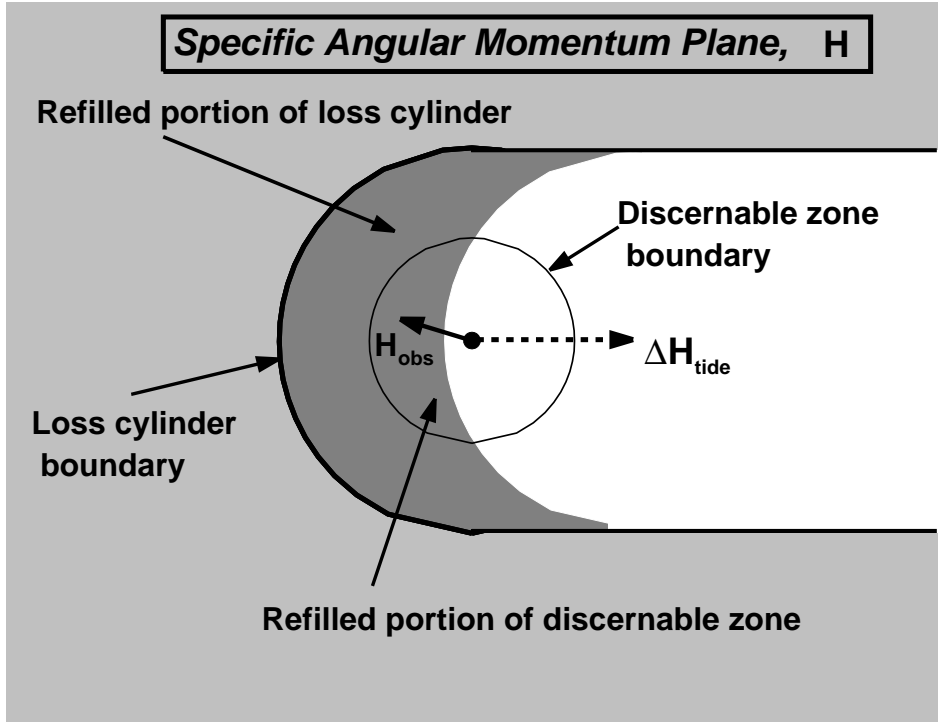} \caption{Schematic illustration of the
plane of specific angular momentum phase space, $\mathbf{H}$, for a given $A$ and $\mathbf{\widehat Q}$, the
aphelion direction defining the normal to this plane. Phase space points illustrated have the {\em same} values of
$A$, $B$, and $L$, but cover the entire region of $\mathbf{H}$ just outside the loss cylinder. Light gray shaded
region outside the loss cylinder boundary ($H > H_{lc}\propto \sqrt{q_{lc}}$) denotes the filled phase space of
comets leaving the planetary region on their prior orbits. The uniform displacement of all phase space points one
orbit later by a tidal perturbation $\mathbf{ \Delta H_{tide}}$ is shown partially refilling the loss cylinder
region (indicated as the dark gray region). The discernable zone within the loss cylinder ($H <
H_{discernable}\propto \sqrt{q_{discernable}}$) is indicated. It remains unfilled for negligible perturbations of
small-$A$ comets, becomes totally refilled for strong perturbations of large-$A$ OOC  comets, and, as illustrated
here, is partially refilled for weak perturbations of intermediate-$A$ OOC  comets. The tidal characteristic
$S\equiv {\rm Sign}(\mathbf{H_{obs}}\cdot \Delta \mathbf{H_{tide}})=-1$  for all {\em observed} comets in the
refilled portion of the discernable phase space here. Over successive orbits the loss cylinder and its shadow
region  of unfilled phase space (shaded white) will be depleted for small-$A$ and intermediate-$A$ OOC comets.}
\end{figure}

A graphical illustration of this theme in Fig. 1 shows the phase
space changes in $\mathbf{  H}$ for a specific choice of
$\mathbf{\widehat Q}$ and $A$ over the course of a single orbit
period. As comets recede from the planetary region on their
prior orbits, the interior of the loss cylinder phase space
region is essentially emptied of OOC comets by planetary
perturbations. The adjacent exterior region remains uniformly
populated in this model. To lowest order in $q/A$, {\em i.e.},
the near-parabolic phase space region just outside the loss
cylinder boundary, the vector displacement in specific angular
momentum, $\mathbf{\Delta H_{tide}}$,  is
independent of the prior value of angular momentum, $\mathbf{
H_{prior}}$, and depends only on $A$ and the major axis
orientation, $\mathbf{\widehat Q}$, which are taken to be fixed
for the phase space of Fig. 1. In a single orbit all nearby
specific angular momentum phase space points, filled and empty,
are displaced uniformly (\cite{ml04}).  Figure 1 illustrates that in the present
modeling, the phrase ``loss cylinder" might be more appropriately
changed to ``loss circle". It also indicates that the phrase
``loss cone" which is commonly used in describing impulsive
domination is no longer appropriate if the tide dominates.

The magnitude of the single-orbit angular momentum displacement is
strongly dependent on $A$, varying as $A^{7/2}$. Small-$A$ comets
are defined here as having negligible galactic perturbations, $<
{|\Delta \mathbf{H}|}_{min}$, unable to repopulate the
discernable zone with {\em new} comets. This includes the inner
Oort cloud (IOC), but also overlaps with the smallest-$A$
population of the formal OOC. Large-$A$ OOC  comets are defined
here as those having strong tidal perturbations
resulting in large displacements in $\mathbf{ H}$ that completely
refill the discernable zone, making $S=\mp 1$ equally likely.
Intermediate-$A$ OOC  comets are defined here as those that are
weakly perturbed by the tide and only partially refill the
discernable zone. Intermediate-$A$ comets preferentially have
$S=-1$, as seen in Fig. 1. In this context intermediate-$A$ OOC
comets have the smallest {\em observed} values of $A$ among the
OOC comets if the tide alone makes the comets discernable. The
fuzzy boundaries between small-$A$, intermediate-$A$ and
large-$A$ depend weakly on $B$ and $L$. We choose the boundaries for these intervals based on data discussed in Section 3.

Therefore, {\em independent of stellar influences on the
 \textrm{in situ} distribution of semimajor axes}, if the galactic tidal
interaction with the OOC dominates impulsive interactions in
making OOC comets discernable at the present epoch we should see
\begin{itemize}
\item a preponderance of OOC comets with $S=-1$ over those with $S=+1$,
and
\item an  association in which $S=-1$
correlates with the smallest observed values of $A$ for OOC comets.
\end{itemize}
Conversely, if  perturber impulses dominate in making OOC comets
discernable at the present epoch, the unique tidal characteristic
$S$ should be a random variable and should be uncorrelated with
$A$.

\subsection{Observational evidence for tidal dominance at the present epoch}
Our data are taken from the $17th$ Catalogue of Cometary Orbits
from which we  convert the ecliptic Eulerian
orbital angles into the galactic angles $B$, $L$ and $\alpha$, the orientation angle of $\mathbf{H}$ defined in Matese and Lissauer (2004).
The Catalogue lists 102 OOC comets (see Appendix) of the highest
quality class, 1A (we count the split comet C/1996-J1A(B) as a
single comet). The quality class predominantly distinguishes the
accuracy of the original semimajor axis determination
(\cite{mse78}). Since our analysis depends sensitively on an
accurate determination of $A$, we restrict our detailed
discussions to class 1A comets. In a previous analysis
(\cite{mww99}) the orbital elements of 82 OOC comets of quality
classes 1A+1B given in the 11{\em  th} Catalogue were used, 47 of
which were class 1A.


\begin{figure}[!hb]
\includegraphics[width=.75\textwidth]{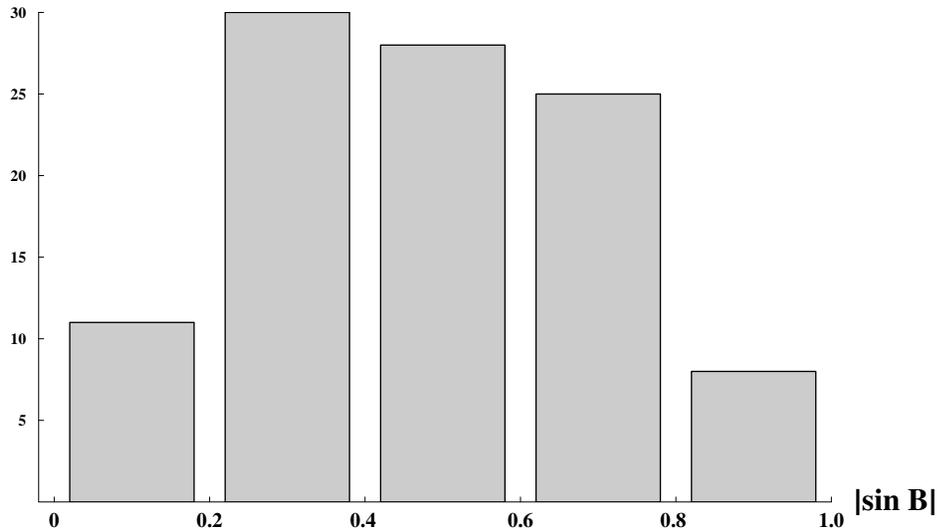} \caption{The distribution in $|\sin B|$ for class 1A new
comets. } \end{figure}

The first observational indication that the galactic tide
dominates involved the distribution in the galactic latitude of
aphelion, $B$ (\cite{del87}). One finds (\cite{mww99}) that the dominant
disk tidal term in the perturbation is $\propto |\sin B \cos B |$.
(Rickman {\em et al.} (2008) conclude that this perturbation is
$\propto |\sin B|$ rather than $\propto |\sin B \cos B |$). $B$ and $L$
are sensibly constant in the course of a single orbit since $\mathbf{
Q}$ has significant inertia for near-parabolic orbits. Therefore,
if the tide dominates we should expect deficiencies of major axis
orientations along the galactic poles and the galactic equator,
and peaks near $B=\pm 45^\circ$. One might argue that
observational selection effects can artificially produce an
equatorial gap, but  polar gaps will be more difficult to
attribute to an observational selection effect.

In Fig. 2 we show the results  presented as a distribution in
$|\sin B|$, which would be uniform for a random distribution.
Polar and equatorial gaps are clear, as predicted if the galactic
tide dominates. A small (but potentially informative) discrepancy
is the location of the peak.
If the tide dominates, our modeling (\cite{ml04}) predicts a peak
at $|\sin B| \approx  0.7$, somewhat larger than seen in the data.
We now look to the tidal characteristic $S$ to further emphasize
tidal dominance.


\begin{figure}[!hb]
\includegraphics[width=.75\textwidth]{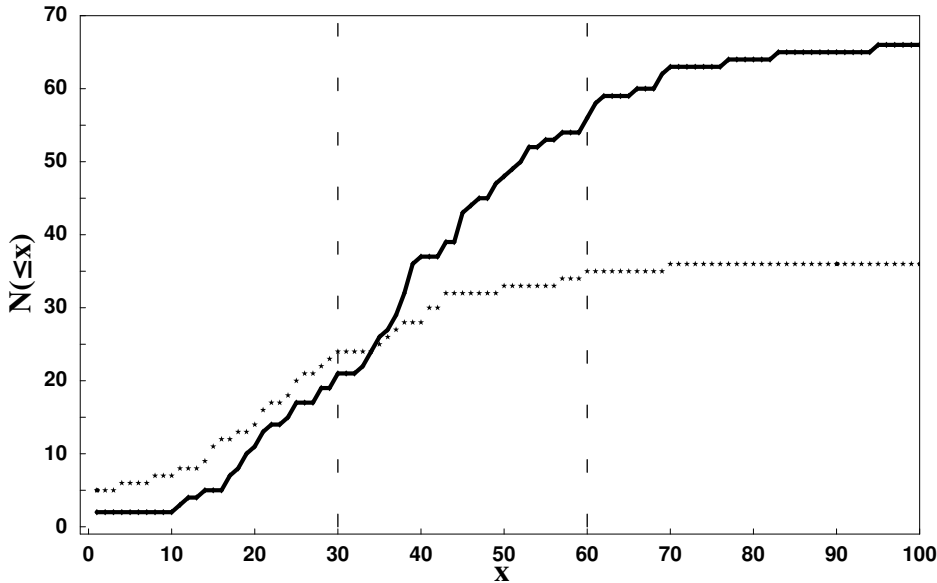}
\caption{The cumulative binding energy  distribution  ($x\equiv
10^6$ AU/$A$) separately illustrated for $S=\mp 1$.  Solid line
$\leftrightarrow$ $S=-1$, dotted line $\leftrightarrow$  $S=+1$. }
\end{figure}

The prediction that $S=\mp1$ should be equally likely for
large-$A$ OOC comets, and that there should be a preponderance of
$S=-1$ for intermediate-$A$ OOC comets (see Fig. 1) is now
considered.  In terms of the original orbital binding energy
parameter $x\equiv 10^6$ AU/$A$, class 1A OOC comets have a mean
formal error of $\approx$ 5 units, but the uncertainty due to
unmodeled outgassing effects is likely to be somewhat larger
(\cite{kres92}, \cite{krol06}). In particular the number of
nominally unbound original orbits is likely indicative of the
true errors embodied in the tails of the class 1A OOC energy
distribution.

In Fig. 3 we show the cumulative class 1A binding energy
distribution of  66 comets with $S=-1$ and 36 comets with $S=+1$.
The binomial probability that as many or more would exhibit this
imbalance if in fact $S=\mp 1$ were equally likely is $2 \times
10^{-3}$. Further, as predicted by tidal dynamics, this
preponderance of $S=-1$ also correlates with intermediate-$A$ OOC
comets in a statistically significant manner. $S=\mp 1$ is
approximately equally likely for  comets with $x \leq 30$,
suggesting that for this range of semimajor axes the galactic
tide is strong enough to refill the discernable zone almost
completely, {\em i. e.}, the tidal efficiency is nearly $100\%$
(\cite{ml02}). We therefore take this as the defining boundary between large-$A$ and intermediate-$A$ comets. In contrast, for $30< x$ the evidence is
that the tide weakens dramatically since 45 comets have $S=-1$
and only 12 have $S=+1$. The binomial probability that as many or
more would exhibit this imbalance if in fact $S=\mp 1$ were
equally likely for this energy range is  $7\times 10^{-6}$.  This
is unambiguous evidence that the class 1A data are of
sufficiently high quality and sufficiently free of observational
selection effects to detect this unique imprint of tidal
dominance in making OOC comets discernable at the present epoch.

Kaib and Quinn (2010) have found that $\approx 5 \%$ of
discernable OOC comets should have $60<x$, the majority of
which have evolved from a primordial location in the IOC. It is
unlikely that these comets have had their prior orbit perihelia
outside the loss cylinder and therefore would not satisfy
 ${|\Delta \mathbf{H}|}>{|\Delta \mathbf{H}|}_{min}$.
Taking account of expected errors in the
binding energies we adopt $30<x<60$ as the defining interval
for intermediate-$A$ comets.

These clear signatures of galactic tidal dominance in making OOC
comets discernable do not mean that we must abandon hope for
detecting any impulsive imprint on the distributions, as we
describe in Sec 3. The characteristic $S=-1$ continues to dominate for observed IOC
comets, suggesting that they are likely to be daughters of OOC comets.
$S=-1$ also remains dominant for other quality class OOC data (1B, 2A, 2B).
However the dominance no longer correlates with the
energy range $30< x $. We interpret this as evidence of
significant errors in the determination of $A$ for these
quality classes, a conclusion made clear by the excess of
nominally unbound comets.

\subsection{Comparisons with recent modeling}
Kaib and Quinn (2009) have done the most complete modeling of the
production of observable long-period comets (LPC) from the Oort
cloud in an attempt to infer the Sun's birth environment. They
closely follow whether an observed OOC comet was born in the OOC
or in the IOC and find that roughly comparable numbers of each
primordial population are made discernable after a suitable time
lapse. Their binding energy distribution for the two discernable
populations (their Fig. S4) differs in that the distribution for
comets born in the IOC peaks at $x\approx37$ while the peak for
comets born in the OOC peaks at $x\approx27$. They conclude that
the IOC pathway provides ``an important, if not dominant, source
of known LPCs". The primordial origin of discernable comets is not
important in our loss-cylinder modeling. The $17th$ Catalogue
indicates that only 2 of the 53 observed class 1A comets with
$100< x <1000$ have future orbits in
the OOC, while 14 of 102 observed class 1A OOC comets that have
future orbits in the OOC. Therefore very few observed comets with
$30<x<60$ are likely to have prior perihelia within the loss
cylinder. We have used their Fig. S4 as a guide to the energy
range most likely to show evidence of a weak impulsive component
of the observed OOC.

Rickman \emph{et al.} (2008) have also embarked on an ambitious
modeling program of the long term evolution of the Oort cloud
which emphasizes a fundamental role for stellar perturbations.
They demonstrate that over long timescales stellar impulses are
needed to replenish the phase space of the OOC which is capable
of being made discernable by the galactic tide. Massive star
impulses serve to efficiently refill this feeding zone for a
period of several 100 Myr and therefore provide one aspect of
the  synergy with the galactic tide that make these comets
discernable.

They also assert that ``treating comet injections from the Oort
cloud in the contemporary Solar System as a result of the
Galactic tide alone is not a viable idea". This statement follows
from their observations that a tide-alone model {\em evolving
over the lifetime of the Solar System} differs significantly at
the present epoch from a combined impulse-tide model.

But their modeling does not shed light on the question of the
dominant dynamical mechanism responsible for injecting OOC comets
into  the inner planetary region {\em at the present epoch}.
Their work makes detailed predictions using results averaged over
170 Myr. Such a large time window inevitably includes many
individual stellar perturbations which contribute directly to
the time-averaged flux. Separating out the dominant perturbation
making these comets discernable at the present epoch would be
best obtained if they (i) ``turn off" both perturbations at the
present epoch, (ii) wait for the discerned flux to dissipate, and
then (iii) alternately determine subsequent distributions
produced by each perturbation separately turned on. This analysis
was not performed.

Indeed, one can compare our predictions (\cite{ml04}) for
discernable distributions in $A$ and the $A$-dependent
discernable zone refill efficiency with Rickman {\em et al.}
(2008) (where they use the term ``filling factor" in the same
context as our ``efficiency"). The tide-alone results {\em at the
beginning} of their simulations (see their Fig. 7 and Table III)
provide the most appropriate comparison since  their phase space
then is most nearly randomized and comparable to that used in our
previous work. One finds that these two sets of distributions are
in good agreement.

Their combined modeling of the latitude distributions does
predict a peak at $|\sin B| \approx 0.5$, more nearly in
agreement with observations shown in Fig. 2 (however they assert
that their predictions do not agree with observations without any
discussion of the nature of the discrepancies perceived). If this
difference with our tidal model prediction of a peak at $|\sin B|
\approx 0.7$ is supported in the future it may provide evidence
that the \emph{in situ} phase space refilling by stellar
perturbations is indeed detectably incomplete at the present
epoch, a point emphasized by them. This would not contradict our
assertion that the unambiguous evidence is that the galactic tide
dominates in making OOC comets discernable at the present epoch.
It remains for them to demonstrate that the modeling adopted here
(\cite{ml02,ml04}) is no longer viable in describing
observations at the present epoch. We conclude our remarks by noting that neither of the above analyses have tracked the orbital characteristic $S$ in the production of
discernable OOC, a major measure of galactic dominance in our work.

\section{Dynamics of a weak impulsive perturbation}
\subsection{Theory}
The dynamics of a weak perturbation of a
near-parabolic comet by a solar companion or field object is now considered. The
change in the comet's specific heliocentric angular momentum induced by a perturber is
given by
\begin{equation}
 \mathbf{\dot{H}_{pert}}=\mu_p
\left( \mathbf{R}\times\mathbf{r} \right)
\left( \frac{1}{|\mathbf{R}-\mathbf{r}|^3}-\frac{1}{r^3} \right)
\label{eqs:dHdt}
\end{equation}
where $ \mu_p={\rm G} M_p$ is the perturber mass located at heliocentric
position $\mathbf{r}$ and $\mathbf{R}$ is the heliocentric comet
position. In terms of the perturber true anomaly, $f$, we have
$
r=p/(1+e_p \cos f),
$
with $p=a(1-{e_p}^2 )$. We then obtain for near-parabolic comets
\begin{equation}
\frac{d \mathbf{H_{pert}}}{df\;\;\;\;\;}=\sqrt{\mu_\odot a} \;
\frac{\mu_p}{\mu_\odot} \;
\left(\mathbf{\widehat{Q}}\times\mathbf{\hat{r}} \right)\;
\frac{R}{b}  \; \left(
\frac{r^3}{|\mathbf{R}-\mathbf{r}|^3}-1\right)
\label{eqs:dHdf}
\end{equation}
where $b=a\sqrt{1-{e_p}^2 }$, the semiminor axis of the
perturber. One then constructs
\begin{equation}
\Delta \mathbf{H_{pert}}=\sqrt{\mu_\odot a} \;
\frac{\mu_p}{\mu_\odot} \; \mathbf{\widehat{Q}}\times \mathbf{I},
\label{eqs:DeltaHpert}
\end{equation}
where the dimensionless integral $\mathbf{I}$ is taken around an interval
 of the perturber true anomaly $\Delta f\equiv f_o -f_i$ that corresponds
 to the comet orbital time interval  between $t_o\equiv 0$, the present epoch, and
 $t_i\equiv  - 2 \pi \sqrt{A^3 / \mu_\odot}$,
\begin{equation}
\mathbf{I}\equiv{\int_{f_i}}^{f_o} \;df \;\mathbf{\hat{r}} \;
\frac{R}{b} \; \left( \frac{r^3}{|\mathbf{R}-\mathbf{r}|^3}-1
\right).
\end{equation}
The first term in $\mathbf{I}$  corresponds to the perturber
interaction with the comet, the second with the Sun. For a
specified perturber orbital ellipse, $\mathbf{I}$ is determined
by the cometary $\mathbf{\widehat{Q}}$ and $A$ as well as the
companion's present value of the true anomaly, $f_o$. Equation
~\ref{eqs:dHdt} is more convenient to use in an impulse
approximation  as has been done in a previous analysis of the
combined tide-stellar impulse interaction with the OOC
(\cite{ml02}). Equation ~\ref{eqs:dHdf} is more appropriate if
one wishes to include the slow ``reflex" effects of a bound
perturber on the Sun.

\subsection{Combined tidal and impulsive interaction}
The galactic tidal perturbation and any putative point source
perturbation of the Sun/OOC are, in nature, superposed in the course
of a cometary orbit. For weak perturbations the two effects can be
superposed in a vector sense,
$
\mathbf{ H_{obs}}  \equiv \mathbf{ H_{prior}}+\mathbf{ \Delta H_{pert}}
+\mathbf{\Delta H_{tide}}\equiv \mathbf{ H_{prior}}+\mathbf{ \Delta H_{net}}.
$
The cometary phase space of comets will have the prior loss
cylinder distribution displaced by $\mathbf{ \Delta H_{net}}$.
The standard step function for the prior distribution of angular
momentum is changed to a uniformly displaced distribution,
similar to that illustrated in Fig. 1, but with $\mathbf{ \Delta
H_{tide}}$ replaced by $\mathbf{ \Delta H_{net}}$,
$
q_{prior} - q_{obs} =  \left( \mathbf{\Delta H_{net}}^2 - 2
\mathbf{ H_{obs}} \cdot \mathbf{\Delta H_{net}}
\right)/2{\mu_\odot}.
$

\subsection{Weak impulsive effects on the discernable energy and
spatial distributions} Matese and Lissauer (2002) modeled the
time-dependent changes in discernable OOC comet orbital element
distributions that resulted from a   {\em weak stellar impulse}.
In particular, the {\em in situ} energy distribution was taken to
be essentially unchanged, as described above. The number of
comets in the large-$A$ ($x\leq 30$) interval that became
discernable after an  impulse was found to be essentially
unchanged from the case with no impulse, although the specific
comets made discernable were changed. That is, the $\approx 100
\%$ efficiency of refilling the discernable zone remains $\approx
100 \%$ throughout the stellar impulse for large-$A$ comets.
This can be visualized in Fig. 1 when we consider a large
$\mathbf{ \Delta H_{tide}}$ that
completely refills the loss cylinder and the comparative case
where the large $\mathbf{ \Delta H_{tide}}$ is modified by a weak
$\mathbf{ \Delta H_{pert}}$. Independent of whether the impulse
slightly increased or decreased the perturbation, a large
$|\mathbf{ \Delta H_{ net}}|$ will still tend to  completely
refill the loss cylinder, albeit with different comets.

For the intermediate-$A$ population  the discussion is more
subtle. Suppose that for a specific $\mathbf{\widehat{Q}}$ a
comet population with $x\approx 35$ will partially fill the
discernable zone due to the tide (see Fig. 1). If some of those
comets experience an impulse that increases $|\mathbf{ \Delta
H_{net}}|$, the discernable zone will be more completely filled.
However other comets will be impulsed such that
 $|\mathbf{ \Delta H_{net}}|$ is {\em decreased}. This
will in turn decrease the number of discernable comets for this
value of $x$ leaving the efficiency for this $x$ only
moderately increased.

Consider now the case for a population with $x\approx 45$. The
tidal perturbation will be smaller by a factor of $\approx 0.4 $
from the $x\approx 35$ population, which may be inadequate to
make any of these comets discernable in our loss cylinder model.
In this case an impulsive torque preferentially opposed to the
tidal torque will have no effect on the number of  observed
comets for this $x$  since none would have been observed in its
absence. But for those comets which have a weak impulsive torque
preferentially aiding the tide, some will be made discernable that
would not have been in the absence of an impulse. Therefore the
efficiency for this $x$  will increase from zero, and it
will preferentially have $S=-1$. The net effect for a weak
impulse is to create an enhanced observed OOC comet population
along the track of the perturber that preferentially has
intermediate-$A$ and $S=-1$. These features have been
demonstrated in detailed modeling for a {\em weak stellar impulse}
(\cite{ml02}). Rickman {\em et al.} (2008) also discuss in detail
this aspect of synergy  but do not consider the importance of the characteristic, $S$, in this discussion.

In reality, the step function distribution of prior orbits for
OOC comets used in the loss cylinder model is a crude first
approximation. A small fraction of original
OOC comet orbits recede from the planetary region as future OOC comets with
perihelia {\em inside} the loss cylinder. This does not
obviate the arguments invoked above. Along with the original
energy errors associated with observational uncertainties and
outgassing effects, we can understand why the observed spread in
original energies seen in Fig. 3 is somewhat larger than
predicted in our loss cylinder model and is consistent with more realistic modeling (\cite{kaib09}).

\section{Observational evidence for an impulse}
\begin{figure}[!hb]
\includegraphics[width=0.75\textwidth]{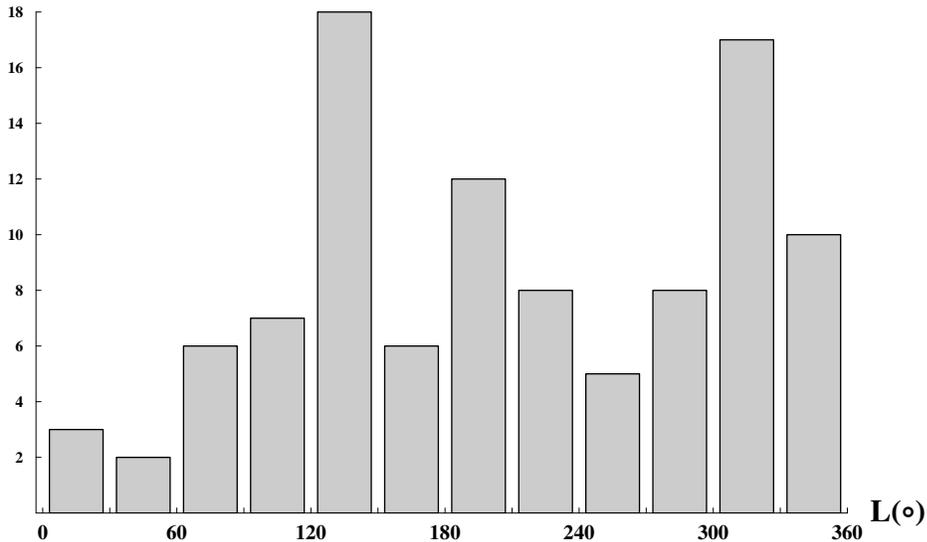}
\caption{The distribution  of aphelia longitude, $L$, for class 1A OOC comets.}
\end{figure}

Matese {\em et al.} (1999) noted an excess of OOC comet major
axes along a great circle roughly centered on the galactic
longitudinal bins $135^{\circ}$ and $315^{\circ}$ using 11{\em
th} Catalogue data of quality classes 1A+1B. Further, they found
that the excess was predominantly in the intermediate-$A$
population and had a larger proportion of $S=-1$, consistent with
it being impulsively produced. In Fig. 4 we now display the
distribution of the aphelia longitudes for our $17th$ Catalogue
data  of quality class 1A. We see that the excess remains in the
present data.

\subsection{Inferring the orbit orientation of a putative bound perturber}
Our first investigation relates to the orientation of the
overpopulated band. There is no obvious  {\em  a priori} reason
why an impulsive track should pass through the galactic poles.
Further, choosing longitude bins for comparison has an obvious
bias in that it preferentially excludes solid angles near the
poles where the tide is weak.

We therefore perform a great circle fit that counts the number of
major axes within an annular band of width $\pm 9.6^{\circ}$,
which has the same solid angle, $4\pi /6$, as the two
overpopulated aphelia longitude bins in Fig. 4. The data that we
now analyze includes the 35 (of 102) $17th$ Catalogue comets
listed in the Appendix which are intermediate-$A$ comets,
$30<x<60$, with $S=-1$.  As described above, this is the subset
most likely to exhibit evidence of a weak impulsive component. Of
these 35 comets, 15 were included in the $11th$ Catalogue
considered by Matese {\em et al} (1999). Counting is done for
{\em all} possible great circle orientations with normal
directions stepped in a grid covering solid angles
$\left({\frac{1}{2}}^\circ \right)^2$. The process is repeated
for the $11th$ Catalogue data, the complete $17th$ Catalogue data
as well as for the 20 comets in the $17th-11th$ Cataogue subset.
Great circle orientations are denoted by the galactic longitude
of ascending node, $\Omega$, and the galactic inclination, $i$. A
similar plot holds for opposite great circle normal vectors
$180^\circ -i$ with $\Omega-180^\circ$.

\begin{figure}[!hb]
\includegraphics[width=1\textwidth]{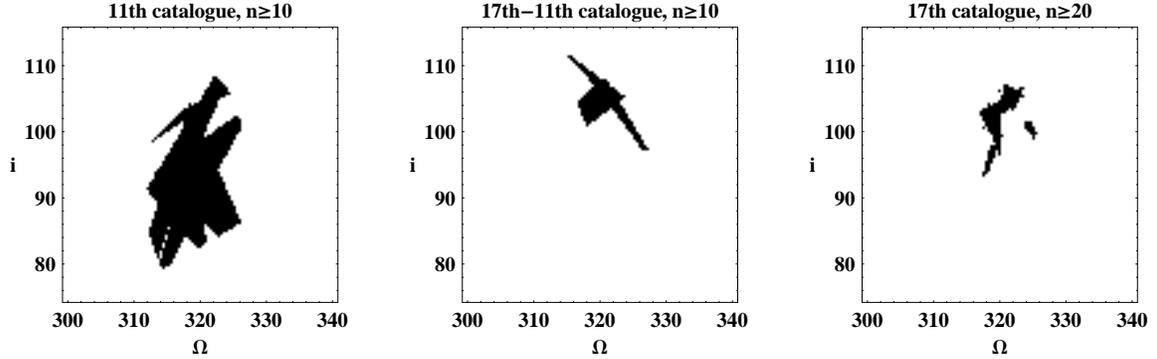}
\caption{Orientation angles ($\Omega$ =galactic longitude of
ascending node, $i$= galactic inclination) of great circle bands
of width $\pm 9.6^{\circ}$ which include 10(20) or more major
axes of intermediate-$A$ comets with $S=-1$ . }
\end{figure}

For the $11th$ Catalogue data  we can find a great circle with as
many as 13 (of 15) axes inside the band. We show all orientations
which include 10-13 major axes in Fig. 5. For the $17th-11th$
data  we show all orientations that include 10-11 (of 20) major
axes. One observes that the peak overpopulated great circle
normal direction occurs in the same general area of the
hemisphere for both the $11th$ and $17th-11th$ data. The
probability that these solid angles would overlap if the two sets
of data were uncorrelated can be estimated to be $\approx 0.02$.
For the $17th$ Catalogue data we show those directions that
include 20-22 (of 35) major axes. The statistical significance of
these observations is discussed below.

The best fit great circle normal direction with a $\pm 9.6^\circ$
band  containing 22 axes of the $17th$ Catalogue data  is
centered at $i=103^\circ,\;\Omega=319^\circ$ (or its opposite
direction). A measure of the uncertainty in the great circle
orientation can be inferred from Fig. 5. In ecliptic coordinates
the best fit great circle plane is specified by $i_e
=133^\circ,\;\Omega_e =190^\circ$, or its opposite direction.

In the Appendix we list the $17th$ Catalogue data for the 102
class 1A comets that constitute our complete set. For each comet
we include the binding energy parameter, $x$, the tidal
characteristic, $S$, the perihelion distance, $q$, and the
galactic angular orbital elements $ B$, $L$ and $\alpha$. The last
column gives the magnitude of the angular separation of the
cometary major axis from the best fit great circle plane,
$\gamma$. Comets having angular separations  within $\pm \gamma$
fall inside a band of solid angle $ 4\pi |\sin \gamma |$. Comets
with year designation prior to 1995 constitute our $11th$
Catalogue data.

\begin{figure}[!hb]
\includegraphics[width=0.75\textwidth]{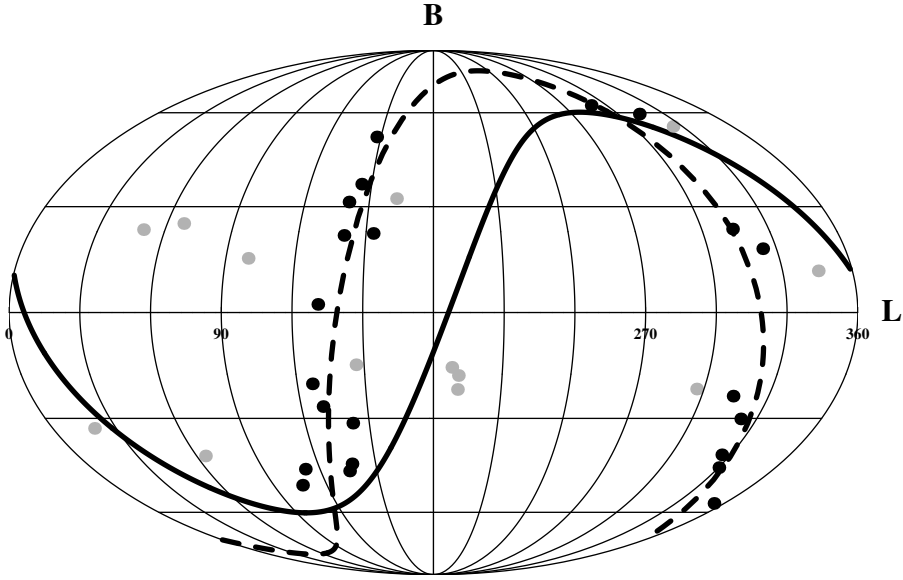}
\caption{The scatter distribution for aphelia directions of
$17th$ Catalogue class 1A OOC comets having a binding energy
parameter, $x\equiv 10^6$ AU/$A$, in the interval $30< x< 60$ and
tidal characteristic $S$=-1. Solid curve: ecliptic plane. Dashed
curve: best fit perturber plane with
$i=103^\circ,\;\Omega=319^\circ$. Black dots: aphelia directions
within a band of width $\pm 9.6^\circ$ covering 1/6 of the
celestial sphere and centered on the best fit great circle path.
Gray dots: exterior to the band.}
\end{figure}


\begin{figure}[!hb]
\includegraphics[width=0.75\textwidth]{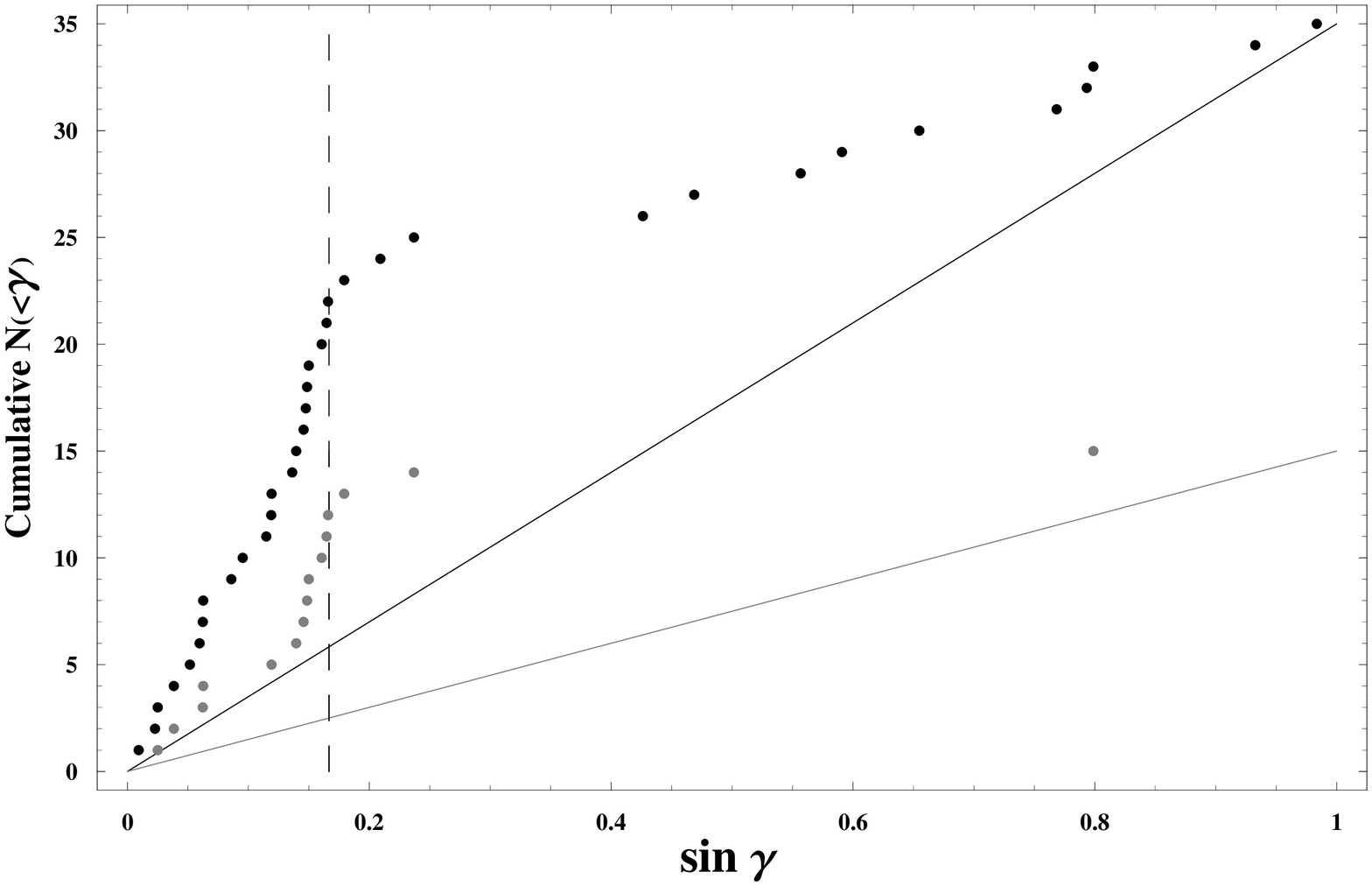}
\caption{The cumulative distribution of the number of major axes
of class 1A OOC comets having $30<x < 60$ and $S=-1$ that fall
within an annular band of solid angle $4 \pi \sin \gamma$
centered on the best fit great circle
$i=103^\circ,\;\Omega=319^\circ$. Black: $17th$ Catalogue data.
Gray: 11{\em  th} Catalogue data. Solid lines are for random
distributions. Dashed line indicates annular band of width $\pm
9.6^\circ $ which includes 1/6 of the celestial sphere.}
\end{figure}

In Fig. 6 we illustrate the scatter of aphelia directions in
galactic coordinates with the best fit great circle arc
($i=103^\circ,\;\Omega=319^\circ$) shown. The data illustrated is
for the $17th$ Catalogue with $30<x<60$ and $S=-1$. Major axes are
distinguished if they are within the $\pm 9.6^\circ$ band. The
distribution of major axes around the great circle arc indicates
comparable numbers in each of the 4 quadrants. This is
inconsistent with an overpopulation produced by a weak stellar
perturbation which produces an overpopulated arc $< 150^\circ$ in
length (\cite{ml02}). In Fig. 7 we show the cumulative number of major axes interior to
the solid angle, $4\pi \sin \gamma$, subtended by a band of width
$\pm \gamma $ centered on the best fit great circle. For
comparison, results are shown for the $17th$ Catalogue data as
well as the corresponding 11{\em th} Catalogue data.

Using an approach that ignores the characteristic  $S$, Murray
(1999) has argued for a solar companion that is the source of a
impulsed observed OOC population with $ 40 \leq x \leq 100$
whose axes are concentrated along a
great circle that is different from that found here.  He found
that {\em all} class 1A comets in the 10{\em th} Cometary
Catalogue  had axes within a $\pm
30^\circ$ band. His point is essentially that all of these
observed comets  {\em must} have been injected into the loss
cylinder by an impulse of a bound companion. We have taken the $17th$
Catalogue data and considered all 1A ($ 40 \leq x \leq 100$)
comets that also were included in Murray's paper which analyzed
$10th$ Catalogue data. A fit to the original data is found to
agree with Murray's result, a best fit companion orbit plane with
a galactic inclination of $i = 37^\circ$ and galactic ascending
node of $\Omega = 175^\circ$  that includes all 12 $10th$
Catalogue axes within a band of $\pm 30^\circ$ which encloses 1/2
of the celestial sphere. When we repeat the fit with the same band
orientation and width we find that 10 of 25 additional comets in
the $17th$ Catalogue are enclosed. The concentration originally
perceived is seen to have disappeared.

\subsection{IR observational limits on the companion's mass and distance}
The absence of detection in the IRAS and 2mass data bases place
limits on the range of possible masses and present distances of
the proposed companion. For  relevant values
the 2mass limits are not very restrictive. Assuming a J-band
detection magnitude limit of 16 and theoretical models of a 4.5
Gyr old 2-10 M$_{\rm J}$ object (\cite{bur03}), we find that the
mass would have to be greater than 7 (10)M$_{\rm J}$ and closer
than $6000 \;(25,000)$ AU for a possible 2mass identification.

For Jovian masses, the strongest current limits on distance
are found from the IRAS Point Source and Faint Source Catalogs.
We have performed a non-parallax search of both IRAS catalogs
using the VizieR Service
(http://vizier.u-strasbg.fr/cgi-bin/VizieR). The search was based
on the theoretical models of Burrows {\it et al.} (2003). Search
criteria included rejection of previously identified point
sources, a PSC 12 $\mu$m flux greater than 1 Jy and a FSC flux
greater than 0.3 Jy. The latter value is based on a completion
limit of 100\% (Moshir {\it et al.} 1992). Only flux quality
class 1 sources were included. We further required that there be
no bright optical or 2mass candidates within the 4$\sigma$ error
limits and that the absolute value of the galactic latitude be
greater than 10 degrees, as required for sources to appear in the
FSC. Distant IR galaxies were excluded by requiring the 25 $\mu$m
flux to be greater than the 60 $\mu$m flux. The PSC search
yielded a single candidate, 07144+5206, with galactic coordinates
B = 25$^{o}$ and L = 165$^{o}$, not far from the great circle
band. The 12 and 25 $\mu$m band flux (0.8 Jy and 0.5 Jy) of this
source is consistent with a $\approx$ 6 M$_{\rm J}$ mass at
$\approx6000$ AU. However, the FSC, which is the definitive
catalog for faint sources, associated this PSC source with
another source located 80 arc sec away. This position
corresponded to an extremely bright (J=6) 2mass source. The
negative IRAS search results suggest that an object of mass 2 (5)
$\rm {M_ J}$ must have a current distance $r \geq 2000\; (10,000)$
AU, respectively.

\subsection{ Dynamical inferences of other perturber properties}
Although the orbit normal of the putative companion is tightly constrained, other properties are less so. If it exists, a solar companion most likely formed in a
wide-binary orbit. The near-uniform distribution of the
overpopulation around the great circle suggests that any putative
companion is likely to have a present orbit that is more nearly
circular than parabolic (${e_p}^2 < 0.5$). The near-circular
implication cannot remain true over the solar system lifetime
since the eccentricity and inclination osculate significantly due
to the tide. Further, the semimajor axis will be affected by
stellar impulses over these timescales.

The implication that ${e_p}^2 < 0.5$, is in fact, consistent with the present
galactic inclination of the perturber inferred here, $i\approx
103^\circ$. This follows from the near-conservation of the $\mathbf{\hat
z}$ component of galactic angular momentum of objects in the IOC
and OOC in the intervals between strong stellar impulses,
$
H_z\propto\ b \;\cos i \approx {\rm constant}. \label{eqs:hz}
$
Since the present value of $|\cos  103^\circ |\approx 0.22$ is at
the low end of a random distribution, $0\leq |\cos i| \leq 1$,
a larger primordial value of $|\cos i|$ implies that
the present value of $e_p$ is reduced from its primordial value.


\begin{figure}[!hb]
\includegraphics[width=0.65\textwidth]{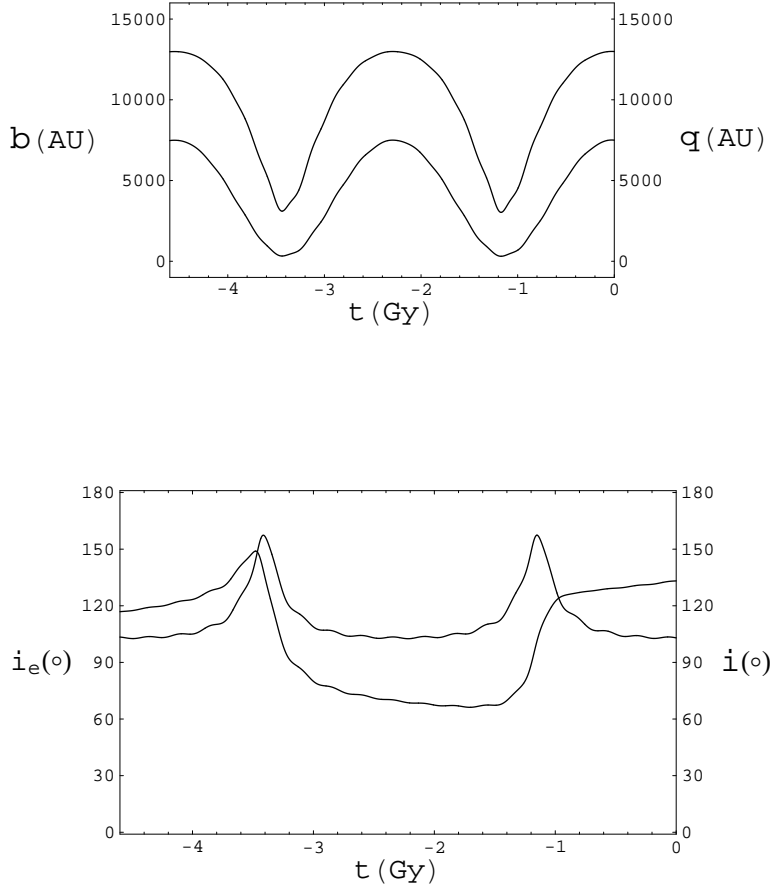}
\caption{Integrating back in
time for 4.6 Gy the osculating orbit of a putative companion
with {\em present} ($t \equiv 0$) galactic orbital elements
$i(0)=103^\circ,\;\Omega(0)=319^\circ,\;\omega(0) =0, \; a=15,000{\rm  AU},\; e_p(0) = 0.5, \;q(0)= 7500$ AU assuming
unchanging galactic environment and neglecting impulsive
perturbations. Shown are the perihelion distance, $q(t)$, and the
semiminor axis, $b(t)$, along with the galactic and ecliptic
inclinations, $i(t)$ and $i_e(t)$.}
\end{figure}

In Fig. 8 we show a representative time dependent companion
orbit  assuming present ($t_o\equiv0$) galactic orbital elements
$i=103^\circ,\;\Omega=319^\circ,\;\omega =0, \; a=15,000{\rm AU}$
and $e_p$  = 0.5. We integrate back in time assuming an
unchanging galactic environment and neglecting impulsive
perturbations, both of which cannot be ignored for more reliable
results. The purpose of the calculation is to demonstrate that
significant changes in the eccentricity and inclination occur
over the solar system lifetime for a companion with $a>10^4$AU.

There is {\em no evidence of overpopulation within the perceived band}
of OOC or IOC comets with $60<x$ where $|\mathbf{ \Delta
H_{tide}}|< |\mathbf{ \Delta H_{min}}|$. Here we must distinguish two subpopulations
of discernable comets with $60<x$ (assuming negligible errors in $x$), (i) comets that
have been directly injected from beyond the loss cylinder by the perturbation  and are making their first orbit as discernable comets, or, (ii) comets that are daughters of $x<60$ comets which had a previous perihelion passage well inside the loss cylinder. The latter is the commonly accepted pathway for producing discernable IOC comets, but requires an {\em ad hoc} imposition of a fading mechanism (\cite{wt99}) to explain the relative numbers of observed IOC and OOC comets.

For a comet major axis
inclined to the perturber orbit plane by an angle $\gamma$
$\;\mathbf{ |\Delta H_{pert}|_{max}}$ varies approximately as $1/
\gamma$, and the cross section for scattering comets varies
approximately as  $\gamma^2$. Thus one expects that a
perturber in a crossing orbit with {\em some}  $60<x$ comets would produce a narrower band of detectable flux perturbed with
$|\mathbf{ \Delta H_{pert}}|> |\mathbf{ \Delta H_{min}}|$.
The absence of such a directly impulsed population (i) is not
inconsistent with the notion that the depletion rate of
this population in crossing orbits within the great
circle band by the perturber would be larger than the rate of
refilling of the band phase space by the galactic tide.

More difficult to explain is the absence of an enhanced daughter population (ii) in the perceived band. A possible explanation is that  the reduced impulse/reflex interaction can remove daughters by a random-walk of their perihelia from inside the discernable zone leading to ejection over one or more subsequent orbits.
A direct numerical integration, including a putative companion, of the type done by Kaib and Quinn (2009) is needed to clarify this matter. This scenario would require $a$ to be at the small end of the allowed range.

Setting the critical perturbation
for $\gamma=9.6^\circ$ equal to the minimum change required to enter the discernable
 zone, Eq. ~\ref{eqs:DeltaHpert} yields
\begin{equation}
 \mathbf{| \Delta H_{pert}|_{crit}}=  \sqrt{\mu_\odot \; a} \;
 \frac{\mu_p}{\mu_\odot}\;\mathbf{|\widehat{Q}\times I|_{crit}}
  \equiv \mathbf{|\Delta H_{min}| }\approx\sqrt{\mu_\odot
  \;5.4{\rm {AU}}}.
\label{eqs:DeltaHpertmax}
\end{equation}
This provides  a means of estimating the perturber properties
needed to weakly impulse the OOC comet population and assist the
tide in making a comet with $30<x<60$
discernable. Here $\mathbf{|\widehat{Q}\times I|_{crit}}$ is
taken to be $ \simeq$ the peak values of
$\mathbf{|\widehat{Q}\times I|}$ sampled over all orientaions of
$\mathbf{\widehat{Q} }$ that are inclined to the great circle by
an angle $\gamma=\pm9.6^\circ$
\begin{equation}
\frac{M_p}{{\rm M_\odot}} \;\sqrt{\frac{a}{5.4 {\rm  AU}}} \;
\mathbf{|\widehat{Q}\times I|_{crit}} =1 . \label{eqs:MpArelation}
\end{equation}


\begin{figure}[!hb]
\includegraphics[width=0.75\textwidth]{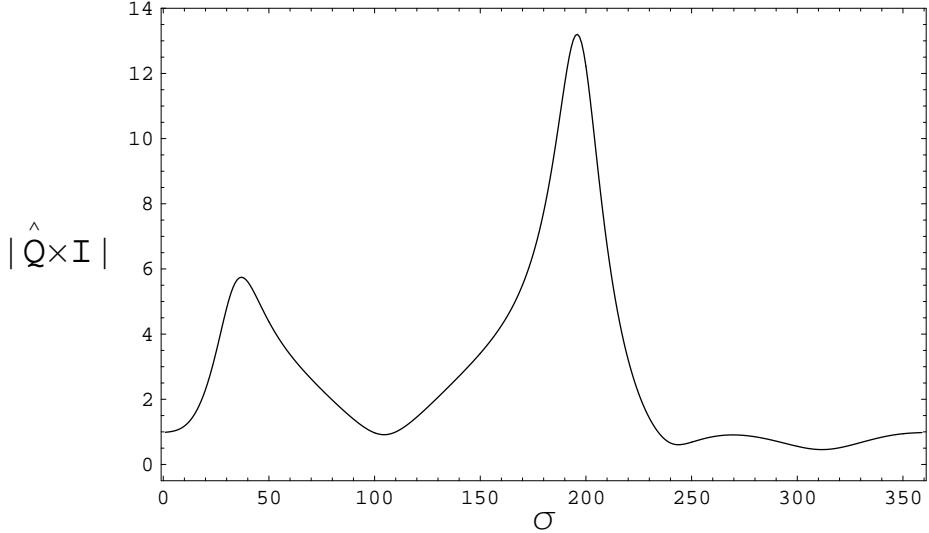}
\caption{An example case showing the single-orbit  impulse
strength $\mathbf{|\widehat{Q}\times I|}$ for comets with
$A=22,000$  AU and   $\mathbf{\widehat{Q}}$ inclined to the perturber
plane by $\gamma=9.6^\circ$. The positions $\sigma=0(180^\circ)$
correspond to comet orientation $\mathbf{\widehat{Q}}$ closest to the perturber
perihelion (aphelion). Perturber parameters are $a= 15,000$
AU, $e_p=0.5$, and $f_o=215^\circ$, the present true anomaly of
the perturber.}
\end{figure}

In Fig. 9 we show an illustrative plot of
$\mathbf{|\widehat{Q}\times I|}$ for perturber parameters $a=15,000{\rm  AU}, \; e_p=0.5,\; f_o=215^\circ$ and comet $A=22,000{\rm  AU}$. A
set of major axis orientations, $\mathbf{\widehat{Q} }$, are
considered which are all inclined along a conical surface at an
angle $\gamma=9.6^\circ$ to the perturber plane. The specific
orientation $\sigma=180^\circ$ corresponds to the case where
$\mathbf{\widehat{Q}}$ is closest to the perturber aphelion
direction. The arbitrarily chosen {\em present} location of the
companion is $f_o=215^\circ$ so that it has recently passed aphelion.
Two peaks are shown corresponding
to the two comet axes $\mathbf{\widehat{Q}}$ that are most
strongly perturbed on the way out and on the way in during the
{\em prior} cometary orbit. The peak at $\sigma=195^\circ$ locates the
perturber position when comets were perturbed on their inward
leg.  The ratio of the comet/perturber periods is $\approx$1.8 in
this case and the second peak at $\sigma=35^\circ$, corresponds
to the outward leg when the perturber was on its previous orbit. The importance of the reflex term can be gauged from the background contribution.
In general the peak corresponding to the inward comet leg will slightly lag the present perturber position if the perturber is presently more nearly at aphelion. The second peak corresponding to the outward leg will, in general, tend to be randomly positioned relative to the inward peak if $A>a$. Each peak contains comet axes $\mathbf{\widehat{Q}}$ that have
angular momentum component impulses that both aid the tide and
oppose the tide.

The results shown are typical for all $e_p<0.5,\;
90^\circ<f_o<270^\circ$ when $10,000{\rm AU}<a<30,000$AU. The
exceptions include the less likely cases when the perturbations on
both the outward and inward legs of a comet orbit find the
perturber at the same general location $f\approx\sigma$ creating
a larger net impulse. We shall adopt a typical value of
$\mathbf{|\widehat{Q}\times I|_{crit}}\approx 6-12$ for all
possible orbit parameters above. Smaller values in this range
typically occur for smaller values of $a$ which have higher
relative velocities at orbit crossing. Another aspect of the variation is that the
small reflex term can add or subtract from the direct impulse
term.

Inserting $M_p=5 {\rm M_J}$ into Eq. ~\ref{eqs:MpArelation} we
obtain $a\leq 6000$ AU. Assuming ${e_p}^2<0.5$, this is
marginally inconsistent with IRAS observational limits. Smaller
masses should not conflict with IRAS. Thus we adopt an
approximate range of perturber parameters $1 {\rm M_J}<M_p<2 {\rm
M_J}$ for $a\approx 30,000\rm {AU}$ extending to $2 {\rm M_J}<M_p<4
{\rm M_J}$ for $a\approx10,000\rm {AU}$. This companion mass range
is consistent with the minimum Jeans mass (1 - 7) M$_{\rm J}$ as
variously calculated (\cite{ws06}, \cite{llb76}).

A solar companion remains a viable option (\cite{mww99,he02}). In addition, we find  that an $\approx 4\;{{\rm
M_J}}$ companion in an orbit such as that shown in Fig. 8 would be capable of adiabatically detaching a scattered
disk EKBO, producing an object with orbital characteristics similar to Sedna   (\cite{mwl05,gml06}). Allowing for
the possibility that the perturber was more tightly bound primordially, smaller masses are then allowed. It may
also be possible to explain the misalignment of the invariable plane with the solar spin axis (\cite{gml06}) if
the putative companion was more tightly bound primordially. Such a companion would have a temperature of $\approx
200$ K (\cite{bur03}) and will be easily seen by the Wide-field Infrared Survey Explorer (\cite{wise}) recently
launched. {\em Gaia}, which will use astrometric microlensing, may also be able to detect the putative companion
and will be launched in 2011 (\cite{gb05}).

Matese and Lissauer (2002) investigated the frequency of weak
stellar showers as well as the patterns of discernable OOC comet
orbital elements. They found that stellar showers  that produce a
$\geq 20 \%$ {\em peak} enhancement in the background tidal flux
occur with a frequency of approximately once every 15 Myr. The
half-maximum duration is $\approx 2$ Myr. Thus it is only
moderately unlikely that we are presently in a weak stellar
shower of this magnitude.

Flux enhancements of this magnitude were found to extend over an
angular arc $\approx 150^\circ$ and have a full-width half-max of
$\approx 50^\circ$. Weak stellar impulses never extend over a
larger arc. More simply put, to get the same stellar-induced flux
enhancement as inferred from the data here, the enhanced region
will have an extent that is $\approx$ half as long and  $\approx$
twice as wide as that observed. Of course one must consider the
possible alignment of two weaker coincident stellar impulses (or
a single weaker statistical anomaly and a coincident single
weaker stellar impulse) that happen to line up on opposite
hemispheres. These will be improbable. For example, suppose we
assume that there will always be two weak stellar impulses
enhancing the observed background tidal flux, each with half the
observed enhancement. The probability that both stellar orbit
planes will be aligned to within $\pm 9.6^{\circ}$ is 0.014.

\subsection{A Monte Carlo test for the likelihood of the conjecture.}
For each of the three data sets discussed above we perform a
variation of the Monte Carlo test done elsewhere  (\cite{he02})
which considered the original conjecture. Each comet in the data
 maintains its measured value of $B$, but is randomly assigned
a value for $L$. The best fit analysis is performed to find the orientation of the
great circle band of width $\pm 9.6^\circ$ which maximizes the
number of major axes interior to the band, just as was done for
the real data. We then record that number, $n$. Repeating the
process for each data set $5000 \times$ we can find the Monte
Carlo probability that $\geq n$ major axes would fall in the band
if, in fact, $L$ was truly random. Results are given in Table 1.
As an example, for the $17th$ Catalogue data, when we randomize
$L$ for the 35 comets we found an unconstrained ({\em i.e.} randomly oriented) band of width
$\pm9.6^\circ$ which maximized the number of major axes contained
in the band at 18 or more  in 30 cases of the $5000$ tested.
\begin{table}[htb]
\vspace{.5 cm}
\begin{center}
{\bf \caption{Monte Carlo probability $^\mathcal{E}
\mathcal{P}_{MC}(\geq n)$ for Catalogue data set $\mathcal{E}$}}
 \vspace{.5 cm}
\begin{tabular}{lrccccr}
\hline
 $\mathcal{E}$  &$n_{max}$&$n\rightarrow$      &  10  & 11   & 12   &     13\\
\hline
$11th$        & 15      &                    & 0.018 &0.0020&0.0002&$<$0.0002 \\
$17th-11th$   & 20      &                    & 0.24  &0.078 &0.015 &0.0014 \\
\hline
              &         &$n\rightarrow$      &18     &19    &20    &21  \\
$17th$        & 35      &                    &0.0060&0.0018&0.0008&$<$0.0002\\
\hline
\end{tabular}
\end{center}
\end{table}

A preliminary discussion of the persistence of the enhancement is
as follows. The Monte Carlo probability,
$^{11th}\mathcal{P}_{MC}(\geq 10)=0.018$ for the original  data
set is superficially significant. The solid angle over which this
occurs can be seen in Fig. 5 to be $< 2\%$ of the celestial
sphere. As many as $n=13$ of 15 axes can be found in a band for
the actual data. Although the Monte Carlo analysis for the
$17th-11th$ Catalogue data set is, in itself, not significant
since at most 11 comets appear in a band for the actual data and
$^{17th-11th}\mathcal{P}_{MC}(\geq 10)=0.24$, the fact that the
cone of normal directions that maximizes the counts for this data
set overlaps with the cone that maximizes the $11th$ Catalogue
data set {\em is significant}. To properly gauge this
significance we discuss a Bayesian analysis of the likelihood of
the present conjecture.

\subsection{Bayesian inference}
The anomalous comet data discussed above can be explained by the
putative companion. We now wish to compare this
hypothesis with other hypotheses that might also explain the
data. The  most obvious alternative explanation is that the
anomaly is entirely a statistical fluke. The other dynamical
explanation is that of a weak stellar shower assisted by the
galactic tide.  Arguments against this possibility have been given
earlier and in the following analysis we assume the probability
of a shower explanation is negligible.

We consider the hypothesis, $\mathcal{H}$, that the there exists a
Jovian mass solar companion orbiting in the Oort cloud with
parameters in the previously discussed ranges that are allowed on
the basis of IRAS/2mass observations. The null hypothesis,
$\mathcal{\neg H}$, is that no such object exists. $\mathcal{E}$
constitutes the evidence considered, the $30<x<60,\;S=-1$ data
sets described in Fig. 5 and Table 1. The \emph{a priori}
relative probability of hypothesis $\mathcal{ H }$, before we
consider evidence $\mathcal{ E }$, is symbolically defined as $
{\mathcal{P}( \mathcal{H} )}/ {\mathcal{P}(\neg\mathcal{ H})}$.

IRAS and 2mass observational constraints prohibit present
locations $r<2000:\;\;10,000:\;\;25,000$ AU for $M_p=2:
\;5:\;10\;{\rm M_J}$ respectively. Dynamical constraints
previously discussed suggest a companion mass in the range
$M_p\approx$ 1 - 4 M$_{\rm J}$ and semimajor axis in the range
$a\approx30,000- 10,000$ AU, respectively. Qualitatively
${\mathcal{P}( \mathcal{H} )}/ {\mathcal{P}(\neg\mathcal{ H})}$
is the ratio of the specified companion parameter space area
($M_p, a$) to the complementary, non-IRAS(2mass)-excluded
parameter space area. Next we try to roughly estimate
this from the non-Catalogue evidence.

We assume that there is a certainty that the sun has an as-yet-to-be-discovered maximum mass companion with mass
between ${\rm M_{Pluto}}\longleftrightarrow 20 {\rm M_J}$, the upper limit being fixed by IRAS/2mass observations.
Zuckerman and Song (2008) give a mass distribution for brown dwarfs down to 13 ${\rm M_J}$ of $dN/dM \propto
M^{-1.2}$. It might be assumed that this power law roughly holds down to the minimum Jeans mass limit of 1 - 4
${\rm M_J}$ (\cite{ws06}) but the mass distribution below 1 ${\rm M_J}$ is completely unknown. For simplicity we
assume $dN/dM \propto M^{-1}$ over the entire mass range.

Next we estimate the probability distribution in semimajor axis
for this companion, assumed to be in the interval $10^2\longleftrightarrow10^5$ AU.
We base this estimate on the known distribution of semimajor axes
of wide binary stars. At least 2/3 of solar type stars in the
field reside in binary or multiple star systems (\cite{dm91}). The
fraction of these systems which have semimajor axes $a\geq$ 1000
AU (wide binaries) is $\approx$ 15\% (\cite{k10}, \cite{dm91} ),
and includes periods up to 30 Myr. The origin of these wide
binaries may be capture during cluster dissolution (\cite{k10}).
For large semimajor axes the observed distribution is well
approximated by $(dN/da) \propto a^{-1}$ and we adopt this power
law for all masses over the entire interval of $a$.

The simplifying assumptions we have made are equivalent to
assuming a uniform probability distribution over the entire
parameter space ($M_p,a$) under consideration
$$\frac{d^2 N}{d\log M_p \; d\log a}={\rm constant}.$$
Thus the combined probability that the sun has a companion of
mass 1 - 4 M$_J$ with semimajor axis 30,000 - 10,000 AU is simply
the ratio of this parameter area  to the entire area after the
IRAS constraint has been utilized. In the following Bayesian analysis we use this value for the {\em a priori} probability ratio, ${\mathcal{P}( \mathcal{H} )}/ {\mathcal{P}(\neg\mathcal{ H})}\approx0.01$.

The likelihood ratio of $\mathcal{H}$ given $\mathcal{E}$ is
\begin{equation}
\Lambda_{\mathcal{H}:\mathcal{E}} \equiv \frac
{\mathcal{P}(\mathcal{E}|\mathcal{H})} {\mathcal{P}(\mathcal{ E
}|\neg \mathcal{H})}
 \label{eqs:likelihood}
\end{equation}
where $\mathcal{P}(\mathcal{E} | \mathcal{H} )$ is  the
conditional probability of evidence $\mathcal{E}$ if hypothesis
$\mathcal{H}$ is true. More meaningful is the \emph{a posteriori}
relative probability, a product of the \emph{a priori}
 relative probability and the likelihood ratio
\begin{equation}
\frac {\mathcal{P}( \mathcal{H} | \mathcal{E})}{
\mathcal{P}(\neg\mathcal{ H}|\mathcal{E})}\equiv
\Lambda_{\mathcal{H}:\mathcal{E}}\;\frac {\mathcal{P}(
\mathcal{H} )}{ \mathcal{P}(\neg \mathcal{ H})}.
 \label{eqs:posteriori}
\end{equation}

\subsubsection{Analysis of the $17th$ Catalogue alone}
In this case we adopt the Monte Carlo results of Table 1 for the
likelihood ratio
$$ \Lambda_{\mathcal{H}:\emph{17th}}=\frac{
1}{{^{17th}}\mathcal{P}_{MC}(\geq n)}$$
 to obtain an \emph{a posteriori} probability ratio
$$
\frac {\mathcal{P}( \mathcal{H} | 17th)}{ \mathcal{P}(\neg
\mathcal{ H}|17th)}=\frac{0.01}{{^{17th}}\mathcal{P}_{MC}(\geq
n)}.
$$
Inserting the results from Table 1 the {\em a posteriori} relative probability ratio is
$\approx 10$ for $n=20$ with corresponding changes
for other values of $n$.

\subsubsection{Separate analyses of the $11th$ and $17th-11th$ subsets}
For $\mathcal{E}=\emph{11th}$ we make the same assumption about
the {\em a priori} relative probability to obtain
$$
 \frac {\mathcal{P}( \mathcal{H} | 11th)}{ \mathcal{P}(\neg\mathcal{
H}|11th)}=\frac{0.01}{{^{11th}}\mathcal{P}_{MC}(\geq n)}
$$
from which we obtain  $\approx 0.5$ for $n=10$
with corresponding changes for other values of $n$.

For $\mathcal{E}=\emph{17th-11th}$ we must take into account that
the \emph{11th} Catalogue data defined a specific cone of
directions in space. In this case we adjust our likelihood ratio
estimate of  the Monte Carlo results of Table 1 by the probability
that the cones of the orbit normals containing maximum counts
would overlap which is $\approx 0.02$ to obtain an \emph{a
posteriori} probability ratio
$$
\frac {\mathcal{P}( \mathcal{H} | 17th-11th)}{ \mathcal{P}(\neg
\mathcal{H}|17th-11th)}= \frac{0.01}{0.02\left[ \;
^{17th-11th}\mathcal{P}_{MC}(\geq n)\right]}
$$
from which we obtain results of $\approx 2$ for $n=10$
with corresponding changes for other values of $n$.

An alternative approach is to allow our $11th$ Catalogue
inference to adjust the {\em a priori} relative probability for
$\mathcal{E} =17th-11th$ ({\em if} the cones overlap) to equal the
{\em a posteriori} result for $\mathcal{E} = 11th$
$$
\frac {\mathcal{P}( \mathcal{H} | 17th-11th)}{ \mathcal{P}(\neg
\mathcal{H}|17th-11th)}=
\frac{0.5}{^{17th-11th}\mathcal{P}_{MC}(\geq n)}
$$
which also yields  $\approx 2$ for $n=10$.

All of these inferences can be scaled if a value of the \emph{a
priori} relative probability is chosen different from 0.01 based
on other evidence than considered here (IRAS/2mass observational
limits and stellar/brown dwarf mass and semimajor axis
distributions). Specifically this value might be increased if one
believed that the Jovian companion hypothesis, $\mathcal{H}$,
provides a plausible explanation for producing Sedna, and it
would be substantially reduced if one believed that the
hypothesis is incompatible with the absence of an enhanced IOC
daughter population along the perceived arc.

We can estimate the size of the impulsive enhancement in the
present data for 102 class 1A comets as follows. Although only 13
of 35 comets lie outside of the $\pm 9.6^\circ$ band for our best
fit, statistically significant results obtain if as many as 17 of
35 were outside the band. So, conservatively, we would expect
$\approx 3$ inside the band if there were no impulsively enhanced
component. The excess of $\approx 15-19$ inside the band is
$\approx 20\% $ of the complementary number of 87-83 for the
complete 1A data  listed in the Appendix. The same number can be
obtained by extrapolating back to zero the large $ \sin \gamma $
portion of the cumulative curve in Fig. 6.

\section{Summary and conclusions}
We have described how the dynamics of a dominant galactic tidal
interaction, weakly aided by an impulsive perturbation, predicts
specific properties for observed distributions of the galactic
orbital elements of outer Oort cloud comets. These subtle
predictions have been found to be manifest in high-quality
observational data at statistically significant levels,
suggesting that the observed OOC comet population  contains an
$\approx 20\%$ impulsively produced excess. The extent of the
enhanced arc is inconsistent with a weak stellar impulse, but is
consistent with a Jovian mass solar companion orbiting in the OOC. A
putative companion with these properties may also be capable
of producing detached Kuiper Belt objects such as Sedna and has
been given the name Tyche. Tyche could have significantly depleted
the inner Oort cloud over the solar system lifetime requiring a
corresponding increase in the inferred primordial Oort cloud
population. A substantive difficulty with the Tyche conjecture is
the absence of a corresponding excess in the presumed IOC
daughter population.

\begin{center} {\bf ACKNOWLEDGMENTS}
\end{center}
The authors thank Jack J. Lissauer for his continuing interest and for his contributions to this research.

\newpage

\begin{center}
\textbf{Appendix}
\end{center}
\begin{table}[htb]
\caption{\hspace{0.5 in} \textbf{Class 1A OOC orbital elements. All angles are in degrees.}}
\begin{center}
\begin{tabular}{lrrrrrrr}
\hline Comet name     &  $x$    &  $S$ &   $q$(AU)  &     $B$
&   $L$     & $\alpha$ &   $\gamma$    \\
\hline
C/1996 J1-A(B) & -510(-1)&  -1  &   1.2978   &    15.9   & 202.6   &55.0  & 51.13  \\
C/1996 E1      &  -42    &   1  &   1.3590   &   -22.6   & 304.3   &319.2  &18.39   \\
C/1942 C2      &  -34    &   1  &   4.1134   &   -24.5   &18.8   &335.2  &  42.27 \\
C/1946 C1      &  -13    &   1  &   1.7241   &   -60.3   &313.1   &276.7  & 14.17  \\
C/2005 B1      &  -13    &   1  &   3.2049   &   -17.6   &82.6   &12.7  &   38.28\\
C/1946 U1      &   -1    &   1  &   2.4077   &    49.0   &347.2   &255.0  & 28.13  \\
C/1997 BA6     &    0    &  -1  &   3.4364   &    27.7   &123.0   &52.9  &  20.05 \\
C/1993 Q1      &    3    &   1  &   0.9673   &    41.2   &112.7   &174.6  & 28.22 \\
C/2005 K1      &    7    &   1  &   3.6929   &   -17.3   &227.4   &312.1  & 85.45  \\
C/2003 T3      &   10    &   1  &   1.4811   &    25.5   &293.6   &148.9  & 16.25  \\
C/2006 L2      &   10    &  -1  &   1.9939   &   -50.0   &225.1   &98.2  &  52.89 \\
C/1974 V1      &   11    &  -1  &   6.0189   &    28.0   &308.9   &63.1  &  2.61 \\
C/1988 B1      &   13    &  -1  &   5.0308   &   -47.6   &316.3   &122.1  & 11.37 \\
C/2006 K1      &   13    &   1  &   4.4255   &    66.1   &56.2   &247.3  &  36.69 \\
C/2004 X3      &   14    &   1  &   4.4023   &   -36.8   &80.9   &24.1  &   31.90\\
C/2005 E2      &   14    &   1  &   1.5196   &    42.0   &312.6   &134.6  & 3.99  \\
C/2006 OF2     &   15    &   1  &   2.4314   &    10.5   &322.5   &133.0  & 5.75  \\
\hline
\end{tabular}
\end{center}
\end{table}
\newpage
\begin{table}[htb]\vspace{-2 cm}\begin{center}
\begin{tabular}{lrrrrrrr}
\hline Comet name     &  $x$    &  $S$ &   $q$(AU)  &     $B$ &
$L$     & $\alpha$ &   $\gamma$    \\
\hline
C/1991 F2      &   16    &  -1  &   1.5177   &    29.7   &90.0   &345.1  & 48.65 \\
C/2005 G1      &   16    &  -1  &   4.9607   &   -55.3   &297.9   &100.6  &22.60  \\
C/1916 G1      &   17    &  -1  &   1.6864   &    17.9   &225.9   &30.1  & 58.95 \\
C/1956 F1-A    &   17    &   1  &   4.4473   &   -25.4   &180.0   &62.1  & 42.33 \\
C/1944 K2      &   18    &  -1  &   2.2259   &   -31.8   &124.3   &159.3  &5.25  \\
C/2007 O1      &   18    &  -1  &   2.8767   &     2.1   &197.3   &7.6  &  55.06\\
C/1935 Q1      &   19    &   1  &   4.0434   &    11.8   &248.0   &115.3  &58.81  \\
C/1999 J2      &   19    &  -1  &   7.1098   &   -49.9   &234.4   &259.3  &52.80 \\
C/1973 E1      &   20    &  -1  &   0.1424   &    17.0&   207.3   &46.3  & 53.15 \\
C/1984 W2      &   20    &   1  &   4.0002   &    33.1   &79.5   &101.0  & 55.69 \\
C/2001 C1      &   20    &   1  &   5.1046   &    -8.3   &135.0   &344.1  &2.02   \\
C/2004 P1      &   20    &  -1  &   6.0141   &    20.2   &208.7   &63.7  & 51.26 \\
C/1922 U1      &   21    &  -1  &   2.2588   &    16.3   &270.6   &52.4  & 39.46 \\
C/1997 A1      &   21    &   1  &   3.1572   &   -19.9   &352.3   &306.7  &25.22  \\
C/2003 K4      &   23    &   1  &   1.0236   &   -46.8   &104.3   &307.2  &12.49  \\
C/2005 Q1      &   23    &  -1  &   6.4084   &    -8.8   &321.7   &258.1  &0.65    \\
C/1947 S1      &   24    &   1  &   0.7481   &    10.9   &185.4   &170.2  &40.61  \\
C/1978 H1      &   24    &   1  &   1.1365   &   -20.7   &124.4   &30.2  & 8.65 \\
C/1999 K5      &   24    &  -1  &   3.2554   &    34.1   &129.8   &271.8  &14.81  \\
C/2001 G1      &   24    &  -1  &   8.2356   &   -46.8   &113.6   &118.9  &7.02   \\
C/2006 HW51    &   25    &   1  &   2.2656   &   -35.4   &162.3   &83.4  & 26.39 \\
C/1914 M1      &   27    &  -1  &   3.7468   &    -1.4   &200.9   &100.9  &59.81  \\
C/1980 E1      &   27    &   1  &   3.3639   &   -18.2   &177.0   &29.9  & 39.78 \\
C/2004 YJ35    &   27    &  -1  &   1.7812   &   -38.2   &346.3   &122.2  &12.23   \\
\hline
\end{tabular}
\end{center}
\end{table}

\newpage
\begin{table}[htb]
\vspace{-2 cm}
\begin{center}
\begin{tabular}{lrrrrrrr}
\hline Comet name     &  $x$    &  $S$ &   $q$(AU)  &     $B$
&   $L$     &
$\alpha$ &   $\gamma$    \\
\hline

C/1992 J1      &   28    &   1  &   3.0070   &   -43.8   &316.9   &0.3  &   10.47  \\
C/1913 Y1      &   29    &  -1  &   1.1045   &   -52.4   &283.0   &237.4  & 31.83    \\
C/1987 W3      &   29    &  -1  &   3.3328   &    64.7   &333.5   &62.5  &  17.92   \\
C/2001 K5      &   29    &   1  &   5.1843   &   -30.3   &223.1   &78.8  &  71.80   \\
C/1989 X1      &   32    &  -1  &   0.3498   &   -41.0   &325.6   &218.0  & 3.57     \\
C/1978 A1      &   33    &  -1  &   5.6064   &   -31.4   &142.3   &167.8  & 9.48     \\
C/1993 K1      &   33    &  -1  &   4.8493   &     2.3   &130.9   &347.4  & 8.38    \\
C/1948 E1      &   34    &   1  &   2.1071   &   -38.6   &250.2   &83.2  &  58.25   \\
C/1948 T1      &   34    &  -1  &   3.2611   &    17.9   &324.0   &306.9  & 8.63     \\
C/2006 Q1      &   34    &  -1  &   2.7636   &   -45.5   &112.2   &134.0  & 8.48     \\
C/1925 F1      &   35    &   1  &   4.1808   &   -40.9   &65.6   &327.8  &  33.96   \\
C/2002 J4      &   35    &  -1  &   3.6338   &    32.4   &162.6   &342.6  & 12.07     \\
C/1974 F1      &   36    &  -1  &   3.0115   &    21.7&   140.3   &10.7  &  3.59    \\
C/2003 S3      &   36    &  -1  &   8.1294   &    11.6   &345.3   &351.9  & 27.95    \\
C/2004 B1      &   36    &   1  &   1.6019   &    27.7   &167.1   &138.4  & 17.55    \\
C/1950 K1      &   37    &  -1  &   2.5723   &   -43.8   &137.7   &269.9  & 8.02     \\
C/1999 U4      &   37    &  -1  &   4.9153   &   -30.2   &322.6   &183.8  & 3.42     \\
C/2006 E1      &   37    &  -1  &   6.0406   &    31.4   &140.5   &290.6  & 21.24     \\
C/2006 P1      &   37    &   1  &   0.1707   &    -5.0   &342.2   &72.9  &  5.47   \\
C/1999 F1      &   38    &  -1  &   5.7869   &    15.2   &99.7   &341.0  &  68.86   \\
C/1999 U1      &   38    &  -1  &   4.1376   &    25.1   &67.4   &291.1  &  40.91   \\
C/2000 CT54    &   38    &  -1  &   3.1561   &    36.7   &145.1   &8.0  &   2.96   \\
C/2002 L9      &   38    &  -1  &   7.0330   &    51.6   &147.6   &292.5  & 4.93     \\
C/1954 Y1      &   39    &  -1  &   4.0769   &   -23.5   &314.1   &232.3  & 9.55    \\
\hline
\end{tabular}
\end{center}
\end{table}

\newpage
\begin{table}[htb]
\vspace{-1 cm}
\begin{center}
\begin{tabular}{lrrrrrrr}
\hline Comet name     &  $x$    &  $S$ &   $q$(AU)  &     $B$
&   $L$     &
$\alpha$ &   $\gamma$    \\
\hline
C/1925 G1      &   40    &   1  &   1.1095   &    10.5   &238.3   &151.2  &   64.77   \\
C/1997 J2      &   40    &   1  &   3.0511   &    -0.5   &261.1   &332.6  &   55.83   \\
C/1954 O2      &   42    &   1  &   3.8699   &   -18.4   &336.4   &53.3  &    11.82   \\
C/1979 M3      &   42    &   1  &   4.6869   &    14.1   &207.0   &126.6  &   55.24   \\
C/2000 K1      &   42    &  -1  &   6.2761   &   -21.6   &190.7   &144.5  &   52.50   \\
C/2006 S3      &   42    &  -1  &   5.1309   &   -15.3   &188.0   &219.4  &   50.21   \\
C/1946 P1      &   44    &  -1  &   1.1361   &   -20.0   &126.6   &108.7  &   6.85    \\
C/1999 Y1      &   44    &  -1  &   3.0912   &    62.7   &289.2   &326.2  &   1.31    \\
C/2000 A1      &   44    &  -1  &   9.7431   &   -33.0   &19.7   &171.9  &    36.22  \\
C/2005 EL173   &   44    &  -1  &   3.8863   &    23.3   &50.4   &350.5  &    79.60  \\
C/1932 M2      &   45    &  -1  &   2.3136   &   -14.6   &146.5   &172.1  &   10.33    \\
C/1987 H1      &   46    &  -1  &   5.4575   &   -17.6   &190.9   &161.4  &   53.02   \\
C/1888 R1      &   48    &  -1  &   1.8149   &    59.5   &313.7   &289.0  &   8.54    \\
C/1983 O1      &   48    &  -1  &   3.3179   &    55.2   &324.4   &355.4  &   13.71   \\
C/1973 A1      &   49    &  -1  &   2.5111   &   -50.7&   106.2   &220.9  &   9.24    \\
C/1989 Y1      &   49    &   1  &   1.5692   &    -6.1   &328.9   &96.2  &    8.22   \\
C/2007 D1      &   50    &  -1  &   8.7936   &   -41.3   &64.4   &157.8  &    33.83  \\
C/2003 WT42    &   51    &  -1  &   5.1909   &   -56.8   &353.7   &171.0  &   6.59   \\
C/2000 O1      &   52    &  -1  &   5.9217   &    23.5   &313.8   &286.2  &   25.23    \\
C/2000 SV74    &   52    &  -1  &   3.5417   &   -21.5   &296.7   &193.5  &   0.53   \\
C/1999 F2      &   54    &  -1  &   4.7188   &   -46.0   &135.3   &112.9  &   6.83    \\
C/1999 S2      &   56    &  -1  &   6.4662   &    22.2   &153.2   &322.9  &   7.83    \\
C/2006 K3      &   56    &   1  &   2.5015   &    46.6   &22.7   &261.9  &    49.79  \\
\hline
\end{tabular}
\end{center}
\end{table}

\newpage
\begin{table}[htb]
\vspace{-1 cm}
\begin{center}
\begin{tabular}{lrrrrrrr}
\hline Comet name     &  $x$    &  $S$ &   $q$(AU)  &     $B$
&   $L$     &
$\alpha$ &   $\gamma$    \\
\hline
C/1976 D2      &   59    &   1  &   6.8807   &   -44.5   &120.9   &315.8  &  3.36    \\
C/1987 F1      &   59    &  -1  &   3.6246   &   -26.5   &129.8   &186.4  &  2.20    \\
C/1993 F1      &   59    &  -1  &   5.9005   &   -45.0   &330.2   &213.7  &  1.44    \\
C/2002 J5      &   60    &  -1  &   5.7268   &   -33.0   &240.2   &157.1  &  67.57   \\
C/2005 L3      &   60    &  -1  &   5.5933   &   -30.3   &203.9   &169.6  &  61.07   \\
C/2000 Y1      &   61    &  -1  &   7.9748   &    28.0   &350.3   &8.5  &    33.52   \\
C/1999 H3      &   65    &  -1  &   3.5009   &   -50.1   &250.1   &167.3  &  49.12   \\
C/1898 L1      &   68    &  -1  &   1.7016   &    28.3   &143.8   &315.4  &  2.01    \\
C/1999 N4      &   68    &  -1  &   5.5047   &   -23.3   &200.6   &204.1  &  61.14   \\
C/1972 L1      &   69    &   1  &   4.2757   &   -40.3   &235.4   &66.6  &   62.09   \\
C/2007 JA21    &   69    &  -1  &   5.3682   &   -31.4   &272.5   &247.5  &  46.07   \\
C/1958 R1      &   76    &  -1  &   1.6282   &   -18.4   &307.6   &133.4  &  14.74   \\
C/1955 G1      &   82    &  -1  &   4.4957   &   -29.8   &233.2   &168.5  &  72.73   \\
C/2005 A1-A    &   94    &  -1  &   0.9069   &    29.4   &113.9   &12.6  &   28.06   \\
\hline
\end{tabular}
\end{center}
\end{table}

\end{document}